\documentclass{ws-mpla}
\usepackage[super]{cite}
\usepackage{graphicx,amsmath,amssymb,slashed,hyperref}
\begin{document}

\markboth{Edwin Ireson}
{Computing Holographic Corrections to the Veneziano Amplitude }

\catchline{}{}{}{}{}

\title{Worldsheets in Holography: \\Computing Corrections to the Veneziano Amplitude}

\author{\footnotesize Edwin Ireson}

\address{William I. Fine Theoretical Physics Institute, University of Minnesota\\
Minneapolis, Minnesota 55455
United States of America\\
eireson@umn.edu}

\maketitle

\pub{Received (Day Month Year)}{Revised (Day Month Year)}

\begin{abstract}
We provide a brief summary of a method to calculate improvements to the Veneziano Amplitude, creating sub-leading non-linearities in the Regge trajectory of states. We formulate it as an extension of a computation by Makeenko and Olesen.  We begin in a confining gauge theory coupled to matter, rewriting the meson scattering amplitude as a specific path integral over shapes and sizes of closed Wilson loops using the worldline formalism. We then prescribe how to further the computation at strong coupling by employing holography, which provides a prescription for the expectation value of these Wilson loops in strongly coupled regimes. 

We find that the problem can then be thought of as a computation in an effective field theory of a string worldsheet sigma model, evolving in a broad class of holographic backgrounds. A convenient interaction picture presents itself naturally in this context, allowing us to draw Feynman diagrams corresponding to the first few corrections due to weaker coupling regimes. The answer we find has qualitatively the same features as other endeavours with the same objective.

\end{abstract}

\ccode{PACS Nos.: include PACS Nos.}
\clearpage
\section{Introduction}	

The Veneziano Amplitude\cite{Veneziano:1968yb} is a cornerstone of String Theory and its uses in particle physics. It was designed in the context of dual models of mesons, which attempted to capture some of the strong-coupling behaviour of bound states in confining gauge theories like QCD. In particular, it was found experimentally that, within a family of mesons and their resonances, these states would have spin proportional to their mass-squared, for reasons that cannot be fathomed from perturbative QCD: $J\propto M^2$. The Veneziano Amplitude's great success was to provide a compact, simple analytical expression that had this linear spin-mass relation, the linear Regge trajectories, baked in as a feature. 

The study of these dual models led to the discovery of what would eventually be called the Nambu-Goto action for a string\cite{DiVecchia:2007vd}, and thereby to String Theory as a whole. While incredibly successful, the original goal of finding a detailed, dual description of QCD using strings still remains tantalizingly out of reach, despite numerous recent advances on the subject. An important example of these is the AdS/CFT correspondence\cite{Maldacena:1997re} and, more generally, the notion of holography: that strong coupling phenomena in $D$-dimensional gauge theories can be translated into weak-coupling, geometric phenomena in string theories evolving in well-chosen curved backgrounds with $D+1$ dimensions. 

While an exact dual model to QCD has not been found, the confinement mechanism within it is generic enough that broad facets of its behaviour are shared by similar theories, including some whose holographic dual are known, enabling us to shed some light indirectly on the subject. One such generic feature of confining gauge theories is of particular interest in this review: indeed, QCD, like many confining gauge theories, experiences RG flow as we move from the infrared to higher energies, and eventually experiencing a deconfining phase transition. In this picture, the Veneziano Amplitude is incomplete: while it captures well the physics of confinement at strong coupling, is insensitive to any RG effect whatsoever, we will show.

Holography, on the other hand, is successful in doing so. It concerns itself chiefly with the study of certain highly curved, asymptotically $AdS$ string backgrounds, endowed with a geometry which purportedly captures the running of the gauge coupling of the theory it is dual to. These spaces have been used in many ways as a means to probe weaker-coupling physics in confined theories, and are particularly useful in modelling the behaviour of Wilson loops of various diameters.

In this work, we review our previous endeavours \cite{Armoni:2015nja}\cite{Armoni:2016nzm}\cite{Armoni:2016llq}\cite{Armoni:2017dcr} in which we propose to use this fantastic feature in order to compute corrections to the Veneziano amplitude due to weakening coupling. First, we review Makeenko and Olesen's Worldline method which constructs the Veneziano amplitude starting from a field theory which is strongly coupled at all energies, thus motivating the need for a correction, and exhibiting exactly where in their proof does their methodology require an improvement. Secondly, we suggest, for very natural reasons, that Holography can be used to perform this improvement and obtain corrections, and explain a number of details and features of holographic backgrounds that are dual to generic confining gauge theories. Thirdly, we will show a worked example of our methodology, using an explicit choice of background to compute a correction to the meson 4-point function, and compare our results to similar endeavours in the past. Lastly, we will show that the correction we find is not tied to the specific choice of background, nor to the details of the 4-point function: the same result will be found for a broad class of confining string backgrounds and identically inside any higher $n-$point function of the theory at hand.

\section{The Veneziano Amplitude from the Worldline}

While the Veneziano Amplitude was first constructed out of (what would become) String Theory, it is a remarkable fact that it appears generically in large-$N$ QCD at strong coupling \cite{Makeenko:2009rf}. The proof of this statement, however, relies on a rather \textit{ad-hoc} Ansatz in order to obtain the result, specifically, an assumption on the behaviour of Wilson loops at strong coupling that is difficult to refine, if only order by order. Namely, if one assumes that every Wilson loop becomes area-behaved (their expectation value is a function of the area of the loop) at strong coupling, various amplitudes one could compute map onto string theory computations. 

\begin{figure}[h]
	\centering
	\includegraphics[width=\textwidth]{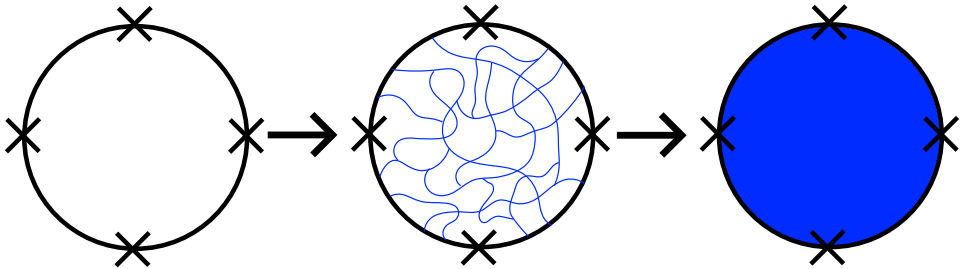}
	\caption{A 4-point meson amplitude progressively fills in at strong coupling to become a tight fishnet, and eventually a disk amplitude}
	\label{fishnet}
\end{figure}

This is not a new idea, the fundamentals relate back all the way to the first applications of dual models of strings such as that of Sakita-Virasoro\cite{Sakita:1970ep}. Pictorially the argument is clear: a tree level amplitude has the topology of a disk. Including perturbative corrections due to gauge interactions fills in the disk more and more, giving it the appearance of a fishnet. At strong coupling, they argue, the sum of perturbative and non-perturbative effects enable the fishnet to close itself completely, and we obtain an amplitude with the topology of a disk, that is, open string scattering amplitudes, as is illustrated by Figure \ref{fishnet}.

We will now outline a derivation of Veneziano-like behaviour in large $N_c$, fixed $N_f$ strong coupling QCD using this idea.

\subsection{The Worldline Formalism}
The Worldline Formalism is a particular way of computing functional determinants in any Path Integral. It is particularly useful in helping disentangle the matter and gauge sectors of QCD-like theories. To wit: in a gauge theory with $N_c$ colours and $N_f$ flavours of massless fermions, the partition function, in terms of the fermion current, reads 

\begin{equation}
	\mathcal{Z}[J]=\int \left[DA\right] \exp(-S_{\rm YM}[A]) \left( {\det} \left( i \slashed{D}+J \right) \right) ^{-N_f}
\end{equation}

This determinant of a functional operator can be evaluated in many ways, here we will employ a geometric approach. We start by the following rewriting:
\begin{align}
\left( {\det} \left( i \slashed{D}+J \right) \right) ^{-N_f}&= \exp\left(-N_f \,\text{Tr} \log\left(\left( i \slashed{D}+J \right)\right) \right) 
\label{expdet}
\end{align}
 This logarithm can then be evaluated with a Feynman parametrisation
\begin{align}
\left( {\det} \left( i \slashed{D}+J \right) \right) ^{-N_f}= \exp\left(-N_f \, \int_0^\infty \frac{dT}{T} \text{Tr}\exp\left(-T\left( i \slashed{D}+J \right)\right)\right)  
\end{align}
The trace is defined over an infinite-dimensional functional space. A convenient way of parametrising the sum over this space is through the introduction of a set of closed loops, all of fixed length $T$ (the Feynman parameter), and sum over all possible shapes of the loop as well as all possible momenta running through it\footnote{This is not the only way to express the determinant in terms of closed  loops, an alternative and equivalent formalism exists which makes use of worldline supersymmetry, of a fermionic coordinate defined on the worldline instead of a momentum variable \cite{Strassler:1992zr}.}. After some algebra, we find
\begin{align}
& \text{Tr}\exp\left(-T\left( i \slashed{D}+J \right)\right) = \nonumber\\
&  \text{Tr}\int \left[Dx\right][Dk]\exp\left(\oint_{\mathcal{C}_T}d{x}^\mu\left(  iA_\mu(x) +J_\mu(x)  \right)\right) \exp\left( i\oint_{\mathcal{C}_T} d\tau \left(  \dot{x}^\mu-\gamma^\mu\right) k_\mu\right) 
\end{align}
where the remaining trace is now only over representation theoretic spaces.

This elegant, if involved procedure introduces a geometric path integral over all possible parametrisations $x(\tau)$ of a closed curve of fixed length $\mathcal{C}_T$, i.e. all possible shapes of such curves. It produces one factor in the integrand which is identifiable with the Wilson operator 
\begin{equation}
\mathcal{W}(\mathcal{C})=\exp\left(  i \oint_{\mathcal{C}} dx\cdot A\right)
\end{equation} for each closed loop at hand. On general group theory grounds the trace of the Wilson operator (in Lie space) runs in inverse powers of $N_c$, thus, at large $N_c$, the exponential introduced in Equation \ref{expdet} can be expanded order by order in $\frac{1}{N_c}$. The first, constant term is vacuous, so we keep only the second term with one Wilson operator in it. Higher orders in this expansion would compute correlations between multiple loops, akin to the genus expansion in string theory (and this similarity is not accidental, as we will argue further). To leading order we choose to focus on the expectation of one loop.

In order to obtain the 4-pion scattering amplitude, we need only differentiate the partition function by the current $J$, since it couples to the pion operator.
\begin{equation}
	\left<\bar{q}q(x_1)\dots\bar{q}q(x_4))\right>=\frac{1}{Z}\frac{\delta}{\delta J(x_1)}\dots\frac{\delta}{\delta J(x_4)} Z[J=0] \
\end{equation}

 Inside the sum over all shapes of loops, this functional derivative has for effect to ``pin'' the loops at $x_1\dots x_4$: the loop can have any shape it likes so long as it passes through those four points eventually. Thus we obtain
\begin{align}
\left<\bar{q}q(x_1)\dots\bar{q}q(x_4))\right>\propto \text{Tr}\int_0^\infty \frac{dT}{T}\int_{x_i \in \mathcal{C}_T} \left[Dx\right][Dk] \left<\mathcal{W}(\mathcal{C}_T)\right>e^{i\oint_{\mathcal{C}_T} d\tau \left(  \dot{x}^\mu-\gamma^\mu k_\mu\right) }
\end{align}
Up to a weight which will soon become irrelevant, the Worldline Formalism expresses our four pion scattering amplitude as a sum of the expectation value of the Wilson operator in pure Yang-Mills, evaluated on a set of curves of all possible lengths and sizes passing through the four prescribed  points.

For practical uses, we would like the above formula in momentum space rather than in position space. The Fourier Transform acts on the worldline: the four external momenta get inserted at four distinct points on the loop, we then sum over all possible points on the loop where this can be done. The closed contours that we are summing over are therefore no longer pinned in place at four points. In detail:
\begin{align}
\left<\prod\limits_{i=1}^4\bar{q}q(p_i)\right>\propto \text{Tr}\int_0^\infty \frac{dT}{T}\int \left[Dx\right][Dk]\prod\limits_{i=1}^4\int d\tau_i e^{i p^\mu_i x_\mu(\tau_i)}
\left<\mathcal{W}(\mathcal{C}_T)\right>e^{i\oint_{\mathcal{C}_T} d\tau \left(  \dot{x}^\mu-\gamma^\mu k_\mu\right) }
\end{align}
It is clear that this computation keeps looking more and more like a string theory calculation: these Fourier kernels are exactly analogous to Vertex Operators, which are used in order to compute the string scattering amplitudes.

\subsection{The Douglas Substitution}

We now wish to take the gauge coupling to be very large and focus on the physics of strongly coupled gluons. The large-$N_c$ approach removes fermion loops from the picture, we now only need to find a way to impose strong coupling at the level of the Wilson loops in the above computation. In their original derivation, Makeenko and Olesen posit the following procedure to materialise the effects of strong coupling: in QCD, very large Wilson loops are expected to show an area-law behaviour, their expectation value should be proportional to the smallest area of all surfaces bound by the loop.
\begin{equation}
	\text{Tr}\left<\mathcal{W}(\mathcal{C}_T)\right>\propto e^{-\sigma \mathcal{A}_{\text{min.}}\left( {\mathcal{C}_T}\right)}
\end{equation}

In order to focus exclusively on this sector of the dynamics, the authors assume that the coupling is so strong that statistically most loops, in the integral over all sizes of loops, has this area-law behaviour. In doing so they push the ultraviolet sector of the gauge theory completely out of the picture and neglects any form of RG flow.

In addition to this, in order for this Ansatz to be anything more than merely symbolic, the authors present a variational problem that produces, as its saddle point, this minimal area for a given loop, a result due to Douglas \cite{Douglas}\footnote{J. Douglas was awarded the very first Fields Medal for precisely this result.}. It is most convenient to formulate it as a path-integral problem
\begin{equation}
	e^{-\sigma \mathcal{A}_{\text{min.}}\left( {\mathcal{C}_T}\right)} = Sp. \int D\theta\, \exp\left(\frac{\sigma}{2\pi}\iint dt_1 dt_2\, \dot{x}(t_1)^\mu \log\left(\theta(t_1)-\theta(t_2) \right) \dot{x}_\mu(t_2) \right) 
\end{equation} 
This path integral sums over all possible reparametrisation functions $\theta$ of the loop. While Douglas' original proof of the statement is written in a very rigorous and analytically-minded way, this statement was rediscovered by Fradkin and Tseytlin\cite{Fradkin:1985qd} and furthered by Polyakov \cite{Polyakov}. They found that the above path integral can be derived from the Polyakov action for an open string with a fixed, externally-imposed boundary. After integrating the interior degrees of freedom, we are left with the above action, reparametrisation invariance forces us to still sum over all reparametrisations of the boundary itself.

Thus, the strong-coupling condition of Makeenko-Olesen is implemented from the worldline formalism by taking the following steps:

\begin{enumerate}
	\item  Every Wilson loop expectation value is replaced by the above reparametrisation path integral, substituting for the notion that all the loops experience strong coupling no matter their size. 
	\item We assume that the string tension $\sigma$ that appears in the above formulae is very large. 
	\item As a consequence, we observe that the remaining weight term in the path integral is negligible.
\end{enumerate}
This second point is a reasonable assumption, on one hand because we require only the saddle point of the Douglas action, i.e. we desire only the ``classical'' configuration of the path integral over $[D\theta]$. This is readily achieved by sending the string tension, which fills the same role as $\hbar^{-1}$ in a QFT Lagrangian, to infinity. On the other hand, this is a generic result from the lattice simulation of QCD flux tubes between the two quarks composing a pion. {This approach is in fact extremely reminiscent of the Strong Coupling Expansion in Lattice Field Theory, while the worldline formalism is connected to the Hopping Expansion. We investigated this avenue\cite{Armoni:2017dcr} with great success, the continuum limit of this pair of Lattice expansions was found to produce this worldline approach.}.  

It is then almost unsurprising that the Veneziano amplitude comes straight out of this set-up, as we have stated, the Douglas action, the core of the computation, is a rewriting of the Nambu-Goto action for a string. After moving to momentum space, both this field theory approach and the open string amplitude produce integrals known as Koba-Nielsen\cite{Koba:1969kh} amplitudes as a result, in the case of four-point functions this subsumes to the Veneziano amplitude

\begin{align}	
	\left<\bar{q}q(p_1)\dots\bar{q}q(p_4))\right> &= \int\,dz_1 dz_2dz_3dz_4 \prod\limits_{k\neq l} |z_k-z_l|^{\alpha^\prime p_k\cdot p_l} \nonumber\\
	&= B(-\alpha^\prime s, -\alpha^\prime t) + (s\leftrightarrow t \leftrightarrow u)
\end{align}

While an incredible feat, this computation motivates the following statement. The Veneziano amplitude is the central feature of the strong coupling regime in confining gauge theories, that is, those theories whose Wilson loops have an expectation value presenting  the area-law, but is insensitive to weaker coupling regimes or any RG flow phenomenon. It is in this spirit that we look for corrections to it. 

These corrections are not automatic, we note. As clever as the set-up we have shown is, the Douglas Ansatz offers no clear algorithm with which to produce corrections. It was inserted by hand in the computation as a well-motivated guess, but, unlike  similar approaches in Lattice Field Theory, it is not the first term in some order-by-order expansion of the action. We need to improve the framework itself. This is precisely why we employ Holography: it is extremely useful in representing in a geometric way the behaviour of Wilson Loops in Field theory.

\section{Holographic Computation of Wilson Loops}

\subsection{Holographic Backgrounds and the Confinement Property}

The salient feature of Holography is the use of highly-curved string backgrounds that are purportedly dual to certain field theories. These string spaces usually extend infinitely in $d+1$ dimensions, i.e. one higher than the field theory it is dual to, along with some compactified extra dimensions, finite in extent, in order to have a consistent 10 dimensional string theory. The extra infinite dimension, and the changes in geometry as one moves up and down it, carry information about the RG behaviour of the field theory. Invariably, in the asymptotic regime of this direction, the spaces in questions will behave more and more like Anti de Sitter space, which has a boundary: the statement of the duality is that the field theory degrees of freedom living on the boundary of the space are dual to string theory degrees of freedom in the bulk of the space.

For our purposes we will phrase the Holographic duality strictly in terms of Wilson loops. The correspondence for this particular object is a lot more intuitive: if we draw a Wilson loop in field theory, the holographic background allows us to compute its expectation value exactly. To do so, we first draw the Wilson loop on the boundary of the space, then create a surface (a string worldsheet), bound by this curve, that protrudes downwards inside the bulk of space. Classically it will settle in a shape that has minimal area as computed in the bulk geometry. The holographic correspondence claims that the total proper area of that surface in the bulk reproduces the behaviour of the Wilson loop expectation value.
\begin{figure}[h]
	\centering
	\includegraphics[width=0.8\textwidth]{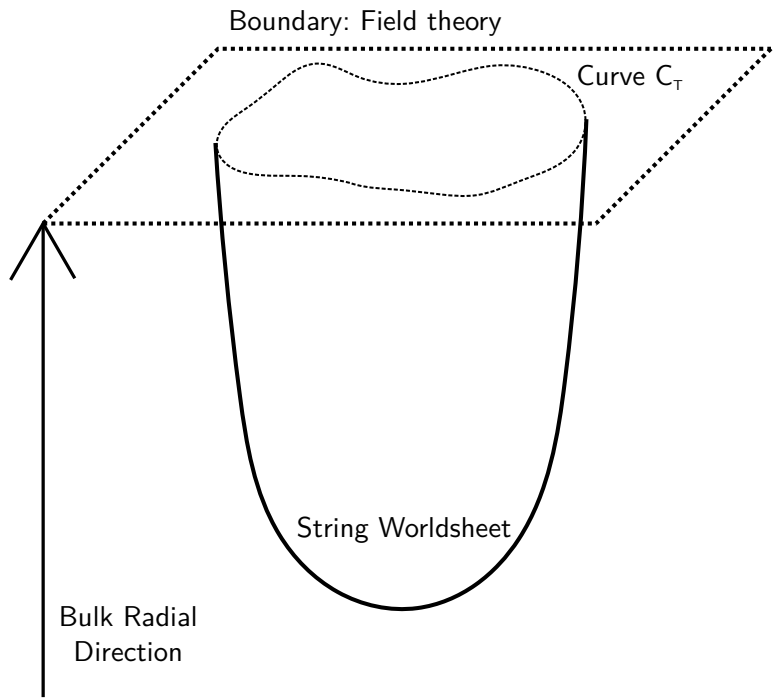}
	\caption{Visual representation of a string worldsheet hanging in the bulk from a Wilson loop drawn on the boundary}
\end{figure}

In the case of $\mathcal{N}=4$ supersymmetric Yang-Mills theory, it is conformally invariant, thus all Wilson loops behave exactly the same and obey a deconfined perimeter-law rather than an area law:
\begin{equation}
	\left<\mathcal{W}(\mathcal{C}_T)\right>\propto e^{-k \mathcal{P}(\mathcal{C}_T)}
\end{equation}

 This theory is dual to $AdS_5\times S_5$, hanging worldsheets within such a space reproduce the above result exactly\cite{Drukker:1999zq}. It may seem bizarre that a surface hanging from a curve has a proper area proportional to that curve's perimeter but it is important to remember that its area is measured in a highly warped space, in this case Anti-de Sitter space.
 
On the other hand, the spaces dual to confining theories exhibit a whole range of behaviours as one moves from the IR to the UV. At high energies, we are past deconfinement, the space looks like $AdS$ space and we approach the UV fixed point. At low energies, we expect to see confinement signalled by an area-law: this is orchestrated by a particular surface in the very deepest parts of the bulk, a flat IR surface on which string worldsheets tend to accumulate most of their proper area.

Dual spaces that exhibit area-behaved string worldsheets have stringent restrictions on them: Kinar, Schreiber and Sonnenschein derived a theorem\cite{Kinar:1998vq} stating sufficient requirements for the space for this phenomenon to take place. Assume an infinite, translation-invariant Wilson loop along the time direction $t$, label the other (finite) spatial direction $x$ and the holographic bulk coordinate $U$. Then, the shape of the Wilson loop is prescribed by the function $U(x)$ (or $x(U)$, but metrics over such spaces typically are written in terms of $U$). Labelling $f^2(U(x)) = G_{00} G_{xx}$ and $g^2(U(x))=G_{00}G_{UU}$, the confinement criterion can be written as follows. Assuming without loss of generality that the boundary of space is at $U\rightarrow \infty$ and the bulk coordinate is bounded from below by $0$: 
\begin{theorem}
 Suppose that  $f,g$ are smooth positive functions over $ 0< U < \infty$, and at $0$ assume the following form
\begin{equation}
f(U)=f(0) + U^k a_k + O(U^{k+1})\,\,\, , \,\,\, g(U)=U^j b_j + O(s^{j+1})
\end{equation}
with $f(0)>0$, $k>0$ a real number, $a_k>0$, $b_j>0$ and $j\geq-1$ also a real number. Also assume that $\int \frac{g}{f^2}$ converges at infinity. \\

Then, $k\geq 2(j+1)$, which also implies that there exists a geodesic line in the space whose endpoints are both on the boundary and whose total energy is proportional to its length. 
\end{theorem}

This can be simplified considerably in the case where both $k$ and $j$ are integers, which is most usually the case for manifolds that are smooth enough. There are then two cases to distinguish:
\begin{enumerate}
	\item If $j=0$, then $k\geq 2$ i.e. both $f,\,g$ are analytic,  and we have an ``end of space'' surface at the bottom.
	\item If $j=-1$ then $g$ is not analytic and $k\geq0$, there is a coordinate singularity at the bottom of the space. This is superficial, however: this surface is the horizon of a Euclidean Black Hole.
\end{enumerate}
In actual fact, even this dichotomy is somewhat illusory, because for practical uses, the singularity of the Euclidean horizon needs to be regularised. In doing so, this maps a space of case $(2)$ into a space of case $(1)$: we will see an example of this in action as we work through a specific case in Section \ref{Witten}. In addition, there is little reason to expect that $k$ is not equal to its lower bound, generically: we expect a full set of non-zero coefficients $a_k$ unless there is a particular feature of the space that prescribes otherwise.

\begin{figure}[h]
	\centering
	\includegraphics[width=0.8\textwidth]{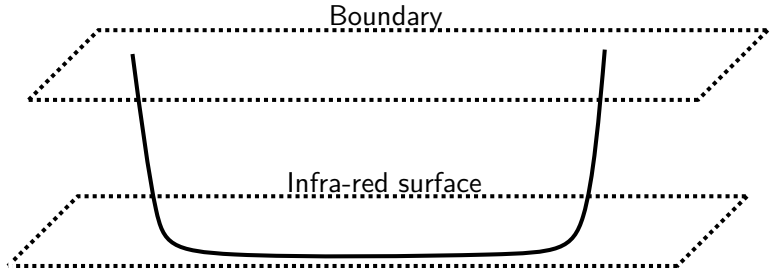}
	\caption{String worldsheet being attracted on an IR surface. The sides contribute negligible proper area, in total one gets an area law.}
\end{figure}

\subsection{Holographically Improved Douglas Ansatz}
We therefore have a clear picture on how to proceed in order to improve the worldline computation: instead of using the Douglas action to represent the behaviour of a collection of Wilson loops, we use the full string action in the background of one of those spaces, and look for parameters we can tune. If we can find a way to tune the space in such a way as to make statistically most loops, even small ones, go down the full depth of space, accumulate on the IR surface, producing area laws at all scales, we will recover the Veneziano amplitude. On top of that, we will have produced it in a flexible framework: we need only undo the process which brought us to the Veneziano amplitude, which forbade any string worldsheet from exploring the rest of the bulk, in order to start incorporating the effects from the weakening of the coupling. We therefore propose the following Holographic Ansatz:

\begin{align}
\left<\bar{q}q(p_1)\dots\bar{q}q(p_4))\right> \propto \int_0^\infty \frac{dT}{T}\int \left[DX\right] \prod\limits_{i=1}^4\int d\tau_i e^{i p^\mu_iX_\mu(\tau_i)}\exp\left( -\sigma\, S_{\text{string}}(G,X)\right) 
\end{align}
where we employ a full (Polyakov) string action in some curved background defined by a prescribed metric $G$. Now, the Fourier kernels truly have turned into Vertex Operators and we are relating the field theory problem directly in terms of a string amplitude.

We will make use of a number of assumptions:
\begin{itemize}
	\item No genus expansion: while the Polyakov action is much more tractable in a path integral, its use leads to the famous genus expansion of string theory. At leading order in large-N, the field theory amplitude we are investigating has no ``windows'', that is, we are taking the expectation value of a single Wilson loop, not a product thereof. This should also be the case for the string amplitude, by the fishnet argument, so in theory $\frac{1}{N_c}$ corrections can be accessed with this method.
	\item Large string tension: $\sigma$ is assumed to be very large (compared to the external momenta, for instance), i.e. $\alpha^\prime = \frac{1}{2\pi \sigma}$ is a small parameter. Lattice verifications of the area law pin the value of the string tension to that of the gauge coupling, therefore, assuming a very strongly coupled regime makes this further assumption consistent.
	\item Supersymmetry: the holographic backgrounds we will use will invariably be spaces of superstrings, since few purely bosonic string backgrounds are practical to use. We have made no mention of this so far, but since we are inserting bosonic Vertex Operators we are more concerned by the bosonic sector of the string theory, at first glance, than the fermionic one. We will provide a more concrete motivation for this when expressing the worldsheet action.
	\item Vertex Operators: strictly speaking, the above equation is not correct. Every string background has its own set of Vertex Operators, of the form $W(\sigma) e^{i p^\mu X_\mu(\sigma)}$ where $W$ can be very intricate (if not even completely unknown), and we are ignoring this fact, to give more generic results. The main purpose of the $W$ operators is to fix matters of spin, they bear Lorentz indices and create the correct spectrum of states in the amplitude (e.g. they remove tachyon poles). We can do both of these things by hand since we know our field theory operator is a massless scalar. Any other feature of the $W$ operators are much more intrinsic to the backgrounds themselves, which, concerned as we are in making generic statements, we find acceptable to ignore.
	\item Compact dimensions of finite non-vanishing size: as explained previously, holographic string backgrounds generically employ compact directions in order to have the full 10 dimensions required for anomaly cancellation. We will typically ignore the influence of those compact subspaces whose extent remains finite and strictly positive all throughout the bulk, as they lead to finite Kaluza-Klein-like corrections which we are not interested in. If this is not the case, typically we will find cone-like substructures inside the space, a compact base that shrinks to zero size at the tip and extends to infinite size upwards. These, as we will mention shortly, play an exceedingly important role. 
\end{itemize}

As a worked example we first provide a simple instance of the steps to follow from this point, using a specific background, in this case Witten's\cite{Witten:1998zw} model of $D4$ branes wrapping a circle.

\section{Witten's model, $D4$-branes wrapping a circle}
\label{Witten}

\subsection{Constructing the Field Theory}
The space at hand is interesting in many ways, but we will not investigate its properties further than the fact it obeys the Confinement Theorem and has a simple enough structure, so as to be comparatively easily treatable by our methods. In addition we are specifically concerned by its geometric properties above all else: we need to provide a metric for our procedure to continue, which this space provides. We will later on argue that the results we find in this specific model are in fact quite generalisable to most spaces obeying the Confinement Theorem.

This space has the following metric: letting $f(U)=1-\frac{U_{\text{KK}}^3}{U^3}$,
\begin{equation}
ds^2=\left(\frac{U}{R}\right)^{3/2} \left(d X^2+\text{d$\tau $}^2 f(U)\right)+\left(\frac{R}{U}\right)^{3/2}\left( \frac{d U^2 }{f(U)}+ U^2d\Omega _4\right) 
\end{equation}
The coordinates are as follows: $U$ is the internal bulk coordinate, at $U=U_{\text{KK}}$ the space has an apparent singularity. At $U\rightarrow \infty$ we find the $(3+1)$D boundary of the space, parametrised by $X^\mu$. In addition there are two compact spaces, a circle $S_1$ described by the variable $\tau$ and a nondescript 4-sphere $S_4$. Importantly, the $S_1$ is the base of a cone which shrinks to zero size at the horizon, while the $S_4$ has finite size at that point. We choose to ignore the latter, though the former will be crucial in setting up a worldsheet theory.

As we have mentioned, the singularity structure of this space enables it to satisfy the Confinement Theorem. However, this is only a coordinate singularity, the space behaves much like a Euclidean Black Hole Horizon around $U=U_{KK}$. As is usual in such cases, we want to regularise this fictitious singular behaviour in order to extract useful information. 

To do so, we perform the following change of coordinates:
\begin{equation}
U=U_{\text{KK}}\left( 1+\frac{u^2}{U_{\text{KK}}}\right)
\end{equation}
 The singularity then vanishes, and we find a more regular space, which still satisfies the Confinement Theorem, this time in the other, regular scenario. However, this was done at the cost of making the new $G_{\tau\tau}$ metric element vanish at the horizon. In detail, letting 
\begin{equation}
\lambda=\left(\frac{U_{\text{KK}}}{R}\right)^{3/2},\, G(u)=\frac{u^2}{U_{KK}^2}\frac{(1+\frac{u^2}{U_{KK}^2})^{3/2}}{(1+\frac{u^2}{U_{KK}^2})^{3}-1},\,A=\frac{3\lambda^{3/2}}{U_{KK}^2},\,B=\frac{4}{3 \lambda ^{3/2}}
\end{equation}
 we find that the $(\tau,u)$ submanifold metric is of the form:
\begin{equation}
ds^2=Au^2\frac{d\tau^2}{G(u)}+BG(u)du^2.
\label{nearflat}
\end{equation}

This is very unfortunate: the metric sees some of its elements vanish around the horizon. This makes the path integration measure singular: in order to maintain reparametrisation invariance, the measure is defined as
\begin{equation}
\label{measure}
[DX]\equiv\prod\limits_{\tau} \sqrt{\det(G(X(\tau)))}DX(\tau)=\left( \prod\limits_{\tau} DX(\tau)\right) e^{\int d^2\tau\,\frac{1}{2}\text{Tr}(\log G(X(\tau))) }
\end{equation} 
This term can be added to the action as an effective potential. Clearly, in either formulation, if the determinant of the metric vanishes, a serious instability develops in the theory. A further change of variables is then required in order to unwrap this cone-like space. A full theory of how to do so in described in Section \ref{genericprocedure}, it exists rather generically and is much akin to the Kruskal-Szekeres change of coordinates, but in Euclidean space, or to a map between conformally warped Polar and Cartesian coordinates. An example of this procedure was shown by Greensite and Olesen\cite{Greensite:1999jw}, deriving the L\"{u}scher term in the QCD string tension using holographic backgrounds.

For simplicity, and since we already want to only include the influence of the near-horizon metric in our computation, we expand all metric elements in powers of $\frac{u}{U_{\text{KK}}}$. We can then observe that we change from pseudo-polar variables $(u,\tau)$ to one complex scalar variable $\Upsilon$ using the following relation
\begin{equation}
	\Upsilon = u e^{iA \tau}\exp\left( \frac{u^2}{4U_{KK}^2}\right) 
\end{equation}

This will work as advertised, as a bonus we notice that the shift symmetry $\tau$ enjoys gets translated into an invariance under changes of phase for $\Upsilon$. In total, we find the following fully regular metric on the worldsheet:

\begin{align}
ds^2=\lambda^{3/2}(1+\frac{3|\Upsilon|^2}{2 U_{KK}^2})dXdX+\frac{4}{3\lambda^{3/2}}\left(1-\frac{|\Upsilon|^2}{U_{KK}^2} \right) d\bar{\Upsilon}d\Upsilon+\dots + O\left( \frac{|\Upsilon|^4}{U_{KK}^4}\right) 
\end{align}
We also check that the determinant is now non-vanishing at $u=0$:
\begin{equation}
\det(G)=U^8_{KK}\frac{16}{9\lambda^{3}}\left(1+6\frac{|\Upsilon|^2}{U^2_{KK} }+\cdots \right) \det\left( S_4 \right) 
\end{equation}

This term can be added to the string worldsheet action: exponentiating its logarithm like we did in Eq.(\ref{measure}) and Taylor-expanding it, this generates a parametrically small mass term for the new radial field:
\begin{equation}
m^2=\frac{9\lambda^{3/2}}{2TU^2_{KK}}
\end{equation}
Note the appearance of an inverse factor of the length scale $U_{\text{KK}}$, this mass is therefore a very small number, this will be relevant later on. 

Finally, there is one last term we need to consider in the action, due to the insertion of Vertex Operators. They can be rewritten as an explicit non-zero current, which sources the $X^\mu$ fields:
\begin{equation}
	\sum_i p^\mu_i X_\mu \left( \tau_i\right)= \int d^2\tau\,  \left( \sum_i p^\mu \delta(\tau - \tau_i)\right) X_\mu(\tau) \equiv \int d^2\tau\,  J_0^\mu(\tau) X_\mu(\tau)
	\label{momcurrent}
\end{equation}
Once all of these operations are performed, the Polyakov action now has the form of a well-defined perturbative field theory, albeit one with an explicit current. In terms of Feynman rules, this current acts as a 1-leg vertex which comes to stop a propagator. Indeed, the Feynman diagrams we will construct to compute the operator expectation value has no ingoing/outgoing definite particle states: none of the diagrams can have external legs in the usual sense. We turned the operator insertions into a term inside the action, thus the diagrams are being generated directly from the partition function at non-zero current, $Z[J_0]$. Therefore, they are all technically vacuum diagrams, but in counting them we will discard true vacuum diagrams that do not depend on the externally imposed momenta, creating this source current. In practice this means that we consider Feynman diagrams with any number of ``external''  legs of the $X$ particle, but sum over all external momenta by Fourier Transforming to worldsheet-position space. To avoid confusion we will distinguish ``inner'' and ``outer'' legs of our diagram, which, \textit{graph-theoretically} are equivalent notions to internal and external legs, but as far as these diagrams relate to integrals and expectation values they are separate notions.

Let us briefly summarise the Feynman rules in this theory: in units where $T=1$,
\begin{itemize}
	\item \includegraphics[width=0.15\textwidth]{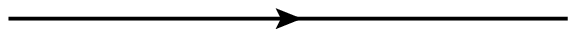} : $X^\mu$ propagator, $\delta^{\mu\nu} \dfrac{1}{p^2}$
	\item \includegraphics[width=0.15\textwidth]{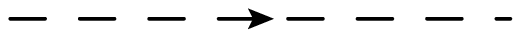} : the $\Upsilon$ propagator, $\delta^{ij}\dfrac{1}{q^2+m^2}$
	\item \includegraphics[width=0.06\textwidth]{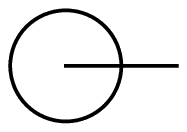} : the 1-leg  $ J_0^\mu(\sigma)$-insertion vertex, $\lambda^{-3/4}\sum\limits_{i=1}^4 k_i^\mu\exp(ip\cdot\sigma_i)$
	\item \includegraphics[width=0.1\textwidth]{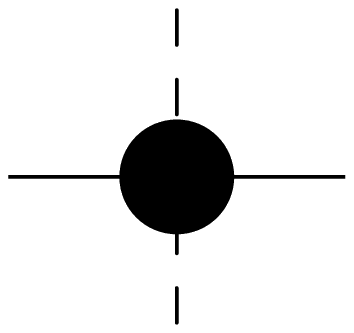} : the 4-leg $X$-interaction vertex: $\delta^{\mu\nu}\delta^{ij}\dfrac{9\lambda^{3/2}}{8U_{KK}^2}(p_i\cdot p_j)$ 
\end{itemize}

There should, in all good measure, also be couplings of these bosonic quantities to various superpartner fermions, since the background space is supersymmetric. However, as one can readily observe, the coupling of the $X$ variable to fermionic quantities happens to vanish for symmetry reasons. Since we must only investigate diagrams with $X$-type ``outer'' legs, there are, therefore, no first-order corrections to investigate that involve any fermions, thus we will ignore them.

Here is an example, the simplest loop diagram that exists in the theory:

\begin{figure}[h]
\centering
\includegraphics[width=0.5\textwidth]{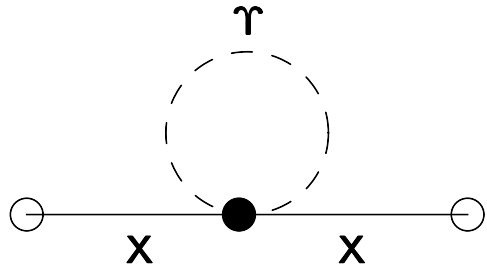}
\caption{First loop correction to the $X$ propagator}
\label{diag1}
\end{figure}

It is a correction to the free $X$-propagator, that is, the diagram which at tree level generates the Veneziano amplitude in our computation. Not only is this diagram divergent, due to the loop momentum, but even the tree-level diagram it corrects is ill-defined: because of the 1-leg vertex, which Fourier transforms diagrams back to worldsheet-position space, we are concerned with having well-defined expressions for position space amplitudes. Unfortunately, even the 2D massless propagator is already ill-defined, but a consistent picture can be obtained by judicious use of regularisation and subtraction schemes. 

It is easy to see that usual dimensional regularisation along with $\bar{MS}$ subtraction is going to be difficult. Every leg is effectively internal, that is, even the outer $X$ legs of the diagrams have their momenta summed over, which leads to more divergences since the tree-level $X$ propagator itself needs a counter-term to be defined correctly. This will quickly lead to an inconvenient amount of book-keeping and redundant computation in order to obtain finite diagrams. However, this operation can be done much more neatly.

\subsection{Regulated Diagrams and Loop Corrections}

We will employ a regularisation and subtraction scheme that is particularly well-adapted to the computation of the idiosyncratic logarithmic divergences in 2D field theories, Analytic Regularisation \cite{Balasin:1992nm}. Much like Dimensional Regularisation, it involves altering the power balance between propagator denominators and integration measure of momenta integrals by an arbitrary parameter. In this case, instead of operating a change on the measure, we alter the power scaling at the level of the denominator. For every propagator in need of regulation in the theory, one proceeds to the following replacement:
\begin{equation}
\frac{1}{p^2}\rightarrow\left[\frac{1}{p^2}\right]= \lim\limits_{x\rightarrow 0}\dfrac{d}{dx}\left( x \mu^{-x}\frac{1}{(p^2)^{1-x}} \right)
\end{equation}
where $\mu$ is an arbitrary mass-scale that exists to keep the overall dimension of the term constant, again like in Dimensional Regularisation. Unlike the latter, this acts on every propagator individually, introducing parameters $x,y,z\dots$  as needed. The fact every leg is acted upon by this procedure solves part of our bookeeping problem already. The other advantage of this method is that it also is a subtraction scheme: the derivative and limit operations do more than isolate a divergence, they actively remove it.  With this prescription, for instance, the position space propagator is automatically finite:
\begin{equation}
\int d^2p\left[\frac{e^{ip\cdot x}}{p^2}\right] =\lim\limits_{x\rightarrow 0}\dfrac{d}{dx}\left( x \mu^{-x}\int d^2p\frac{e^{ip\cdot x}}{(p^2)^{1-x}} \right)=-\frac{1}{2\pi}\log\left(\mu |x| \right) 
\end{equation}

\begin{figure}[h]
	\centering
	\includegraphics[width=0.59\textwidth]{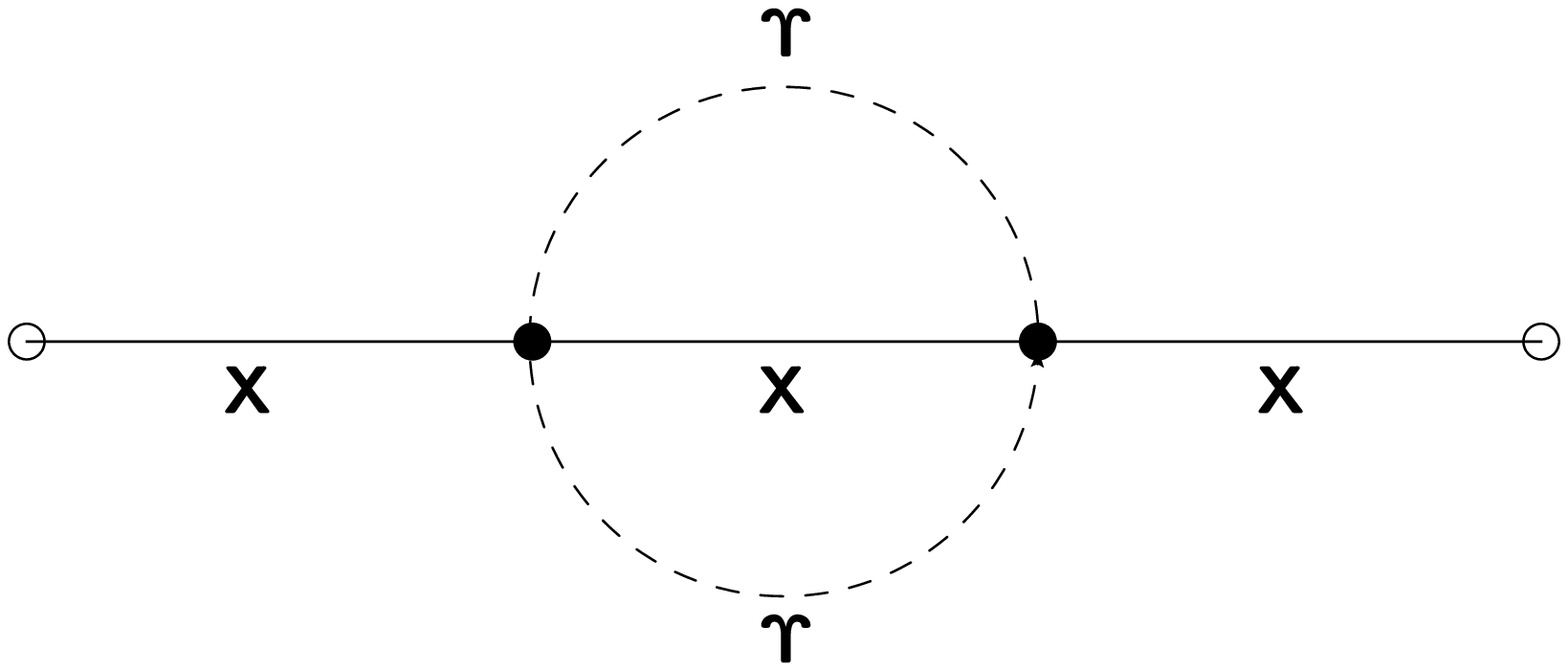}
	\includegraphics[width=0.39\textwidth]{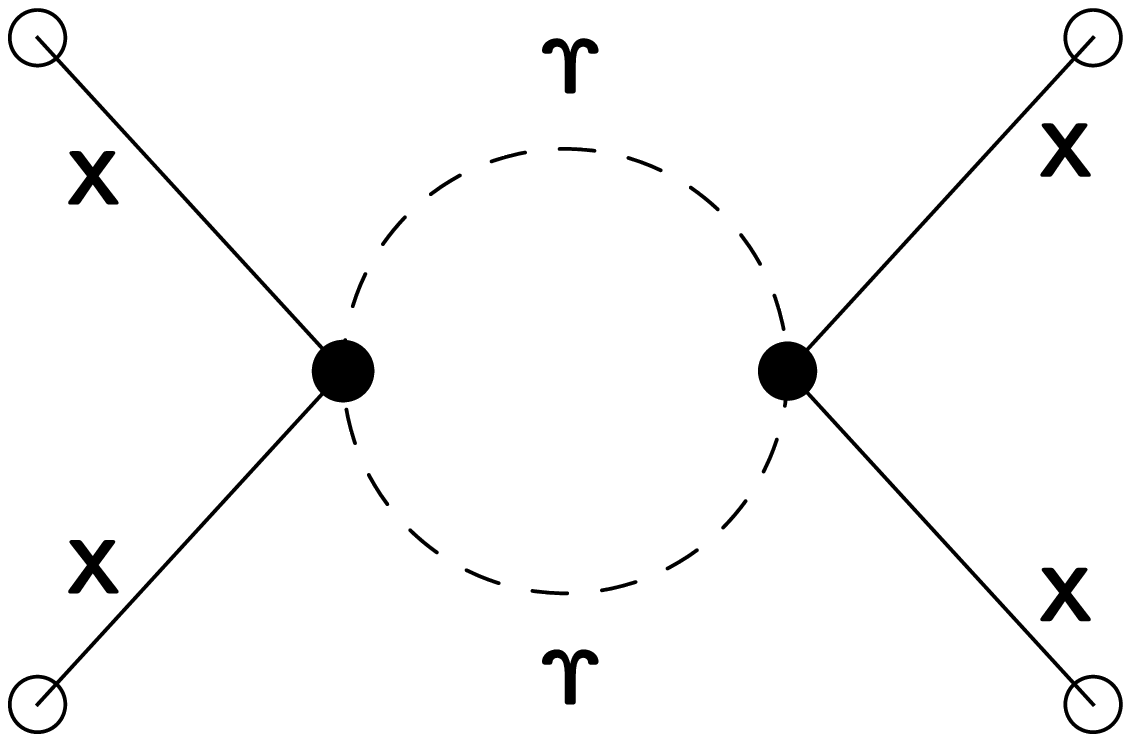}
	\caption{A two-loop diagram of lower order than a one-loop diagram}
	\label{loopdiags}
\end{figure}

This is particularly useful for us, since we expect that every diagram we write will be dimensionless (being ``vacuum'' diagrams in some sense), and so either finite or logarithmically divergent, this method allows us to compute such diagrams in a very systematic manner. In fact, the first diagram shown above in Figure \ref{diag1} is made finite by this procedure: since it has no momentum transfer, its integral structure splits into a ``bubble'' term and a copy of the $X$ propagator. The result is a finite correction to the wavefunction normalisation of $X$, or in other words to the effective string tension of the Veneziano amplitude. We are not interested in such corrections, which will generically happen every time the diagram's integral can be entirely factorised so as to have no momentum transfer at all. Figure \ref{loopdiags} shows the first two ``non-trivial'' diagrams that we can write in this theory. The question is then how to compare them.

At this point we must point out another curious feature of these diagrams: loop order is not a good expansion parameter. Because we are in two dimensions and due to the shape of the 4-leg vertex, which always comes with two units of momenta attached to them, loop number drops out of the equations when computing the superficial degree of divergence of an arbitrary diagram. By inspection of the Feynman rules it appears that the scaling of a diagram with $V$ vertices and $2E$ ``outer'' legs is
\begin{equation}
\left( \lambda\right) ^{3\left(V-E \right)/2 } \left(\frac{1}{U_{\text{KK}}^2} \right)^{V} =  \left(\frac{\lambda^{3/2}}{U_{\text{KK}}^2}\right)^V \left(\lambda^{-3/2} \right) ^{E}
\end{equation}
In order for this to remain under control as we generate diagrams with higher and higher numbers of vertices and edges, we require (in string units) $\lambda^{3/2}\gg U_{\text{KK}}^2\gg 1$. Then, it is clear that loop order truly is meaningless in this theory: it is possible to make certain one-loop diagrams higher order than other, two-loop diagrams, an example is given in Figure \ref{loopdiags}

This is reasonably convenient: this first diagram is more relevant to our interests as it is a correction to the two-point function, the one that produces the Veneziano amplitude at tree-level. Certainly both should contribute, but the two-loop, two-point one contributes first.

Despite its simple appearance, the first diagram of Figure \ref{loopdiags} (sometimes called Sunset or London Transport diagram) is not entirely straightforward to compute, particularly because $\Upsilon$ is a massive field while $X$ is not. Such graphs whose internal symmetry is disturbed by unequal masses lead to notoriously challenging integrals. In four dimensions and with Dimensional Regularisation, much of the ground work for the Sunset diagram with two masses has been arduously done\cite{Smirnov:2012gma}, but neither are at hand here. On top of that, we require the answer in position space, not momentum, which if anything increases the number of integrals to perform. 

We must remember that the mass generated for $\Upsilon$ is parametrically very small. It is in fact of the same order as the coupling of the $XX\Upsilon\Upsilon$ vertex. In addition, the fully massless Sunset diagram is well-defined and finitely regulated by our procedure above. It produces a contribution that is different in form from the Koba-Nielsen integrand usually associated to the Veneziano amplitude. Thus, as a first investigation, we will only look at the massless limit of this diagram.

Despite being finite, well-defined and relatively straightforward, the integral's computation is nonetheless long and verbose. We refer to the fuller text\cite{Armoni:2016llq} for the full derivation, and proceed straight to the result and its interpretation. In total, once we are back in position space, the diagram leads to 

\begin{equation}
I= -\frac{1}{(2\pi)^3}\frac{27\lambda^{3/2}}{64 U_{KK}^4}\left( \sum\limits_{i<j}k_i\cdot k_j  \left(\vphantom\int \log\left((\sigma_i - \sigma_j)^2 m^2\right) \right) ^3 \right)
\label{worldsheetcorrection}
\end{equation}

\subsection{Consequences for the Veneziano Amplitude}

Now that we have computed this quantum correction, we need to insert it in the outer path integral and sum over all insertion positions. The total amplitude will look like the following:

\begin{equation}
\mathcal{A}(k_i)=\delta\left(\sum_i k_i \right) \oint \prod_{i=1}^{4} d \sigma_i \exp\left(-\frac{1}{2}J_0\Delta^{-1}J_0\right)\left( 1 + \dots  \right) 
\end{equation}
Where $J_0$ is the non-zero current issuing from the Vertex Operator insertions, as defined in Eq.(\ref{momcurrent}), and $(\dots)$ imply further corrections. 
 We now introduce the standard notation of Mandelstam variables. Considering every momentum as ingoing, we write
 \begin{equation}
 s= (k_1+k_2)^2 = 2k_1\cdot k_2\,\,,\,\,t=(k_1+k_3)^2=2 k_1\cdot k_3\,\, , \,\, u=(k_1+k_4)^2=2k_1\cdot k_4
 \end{equation}. We will assume that the momenta all sum up to zero always, allowing us to strip off the overall momentum-conserving delta function in the expressions below.
 
The ``tree level'' result, with no corrections, produces the Veneziano amplitude by way of the following Koba-Nielsen\cite{Koba:1969kh} integral:
\begin{equation}
\mathcal{A}(k_i)= \int_{\mathbb{R}^4} \left( \prod_{i=1}^{4} d \sigma_i \right) \left( \prod_{i<j} |\sigma_i -\sigma_j|^{k_i\cdot k_j}\right) 
\end{equation}
Such integrals are ill-defined and infinite, namely because they are $SL(2,\mathbb{R})$ invariant, the physical implications of this fact are well-known in string theory. At any rate, it is usually fixed by mapping three of the points to $0,1,\infty$ leaving one undefined. This then produces the Veneziano amplitude's integral form:

\begin{equation}
\mathcal{A}(k_i)=\int_0^1 dz \left|z\right|^{s}\left|1-z\right|^{ t} +(s\leftrightarrow t \leftrightarrow u)= B(s,t) +(s\leftrightarrow t \leftrightarrow u)
\end{equation}
 
 Now, we can investigate the nature of the first correction: inserting the result of the worldsheet computation given by Eq.(\ref{worldsheetcorrection}), we obtain the following result
 \begin{align}
 \mathcal{A}_1(s,t)= \int_0^1 dz \left|z\right|^{s}\left|1-z\right|^{ t}\left(1+\rho^2 \left( s\log^3\left(|z|^2 \right) +t\log^3\left( \left|1-z\right|^2\right) \right)    \right) 
 \end{align}
 
  The integral, thus modified, can be expressed through the use of the family of functions $\psi^{(n)}$ defined by $\psi^{(0)}=\Gamma^\prime / \Gamma \,\,\, , \psi^{(n+1)}=\psi^{(n)\prime}$. This gives in our case 
 \begin{align}
\mathcal{A}_1(s,t)= &B(s,t)\left(1+ \rho^2\left((\psi ^{(0)}(s)-\psi ^{(0)}(s+t))^3-\psi ^{(2)}(s+t)+\psi ^{(2)}(s)\right.\right.\nonumber\\
 &\left.\left.+3 (\psi ^{(1)}(s)-\psi ^{(1)}(s+t))
 (\psi ^{(0)}(s)-\psi ^{(0)}(s+t))\right) +(s\leftrightarrow t) \right)   
 \end{align}
 where we encapsulate the overall numerical factor in front of Eq.(\ref{worldsheetcorrection}) by $\rho$, which by our assumptions is indeed a small parameter.
 
 We would then like to observe the asymptotics of this corrected amplitude, in order to try and find a new form for the Regge trajectory. For this purpose we write the large $s$, fixed $t$ asymptotic expansion of the above expression, often called the Regge limit of the amplitude. Indeed, one expects that the 4-point amplitude in this limit has the following behaviour
 
 \begin{equation}
 \mathcal{A}^{(4)} \sim\alpha(s)^{\alpha(-t)}
 \end{equation}
 
 We therefore expand the result in this limit and match the expressions to obtain a new, modified form of $\alpha(s)$. We recall the asymptotic behaviour of the $\psi^{(n)}$ functions:
 
 \begin{equation}
 \psi^{(0)}(z) \sim \log(z)\,\,\,,\,\,\, \psi^{(n>0)}(z)\sim z^{-n}
 \end{equation}
 
 Other than $\psi^{(0)}$ all the other functions vanish at infinity, so we expect that the former produces the main contribution. We then get:
 
 \begin{equation}
 \mathcal{A}(s,t)\sim s^{-t}+\rho^2 t \log(s)^3 \sim s^{t(1-\rho^2 \log(s)^2)}
 \end{equation}
 
 This expression then motivates the corrected, non-linear form of the Regge function: 
\begin{equation}
	\alpha(s) = s^{1-\rho^2\log(s)^2}
\end{equation}

\begin{figure}[h]
\centering
\includegraphics[width=0.8\textwidth]{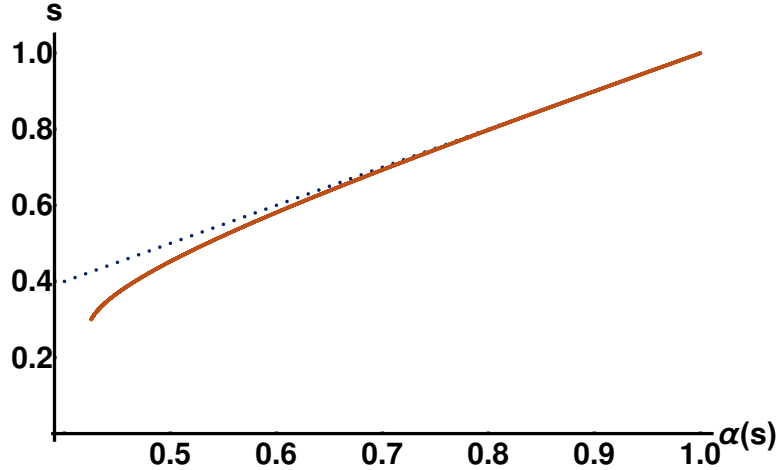}
\caption{The linear Regge function (dashed) and its corrected form (solid) as we have computed it}
\label{plots1}
\end{figure}

How does this new Regge function behave? We provide plots for the uncorrected and corrected function in Figure \ref{plots1}.
We have plotted it only in the regime where the correction is small: going too far down as $s\rightarrow0$ we are bound to go beyond the limits of our approximations. 

\begin{figure}[h]
	\centering
	\includegraphics[width=0.8\textwidth]{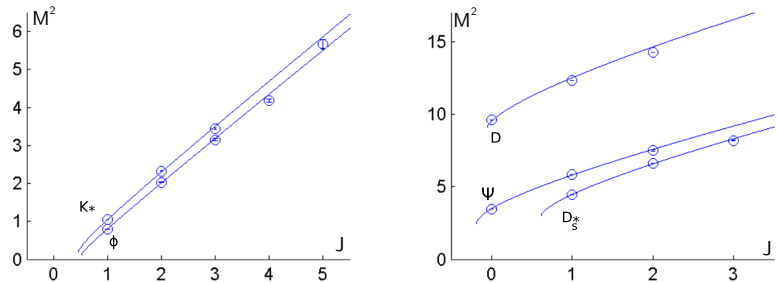}
	\caption{Regge functions bending for several meson resonance trajectories}
	\label{plots2}
\end{figure}

For comparison, a qualitatively similar bending of the Regge function was observed by Sonnenschein \textit{et al}\cite{Sonnenschein:2016pim}, using a spinning string model in a holographic background to find corrections. This model had the feature of having enough undetermined parameters so as to be able to fit it to data issuing from a number of meson resonance families directly from empirical observation, the curves they obtained are shown in Figure \ref{plots2}.
The result of their analysis was that their fit was better than linear regression, providing evidence of this kind of trajectory bending even in usual QCD, which is not at large $N_c$. A similar deviation was also found in an analogous work by Imoto and Sugimoto\cite{Imoto:2010ef}.

Qualitatively, the fitted phenomenon described by Sonnenschein and ours are very similar. While both approaches are based on the concept of probing a certain class of confining holographic string backgrounds, they come to their conclusions via wildly different ways. This helps to convince us of the relevance of the effect we are detecting.

In addition to direct comparison with previous results, it is important to test the sturdiness of our computation by verifying how universal the correction we compute actually is.

\section{Universality of the Correction}

There are two main sources of universality for our correction: one, we must check that it is possible to vary which holographic background we pick at the start of the procedure, and obtain an identical result, two, we can check that the correction is identical in higher-point amplitudes. Since we have hypothesised that it is a correction to the Regge function itself, it should appear quite identically in the multi-Regge limit of higher-point amplitudes. We will now show that both of these facts are true.

\subsection{Generic description of the flattening procedure}
\label{genericprocedure}

We now assume instead a totally generic string background, which only needs to obey the Confinement Theorem. As a result it has an end of space or a horizon, and we wish to make the metric fully regular around that point: two main obstacles can occur, either the metric elements can diverge or vanish in that region, we saw examples of both of these scenarios. The former stops us from writing worldsheet kinetic terms, the latter generates unstable potentials for fields on the worldsheet through non-perturbative terms generated by the path integration measure.

If the metric diverges, the space has the structure of a Euclidean Black Hole horizon in order to have the confinement property, and this horizon is precisely the surface that accumulates worldsheets. This was the case in the example we investigated.

If we want to expand worldsheet fluctuations around that point, we need to change to different coordinates. Fortunately, this is easily done: we know $G_{UU}\sim \frac{1}{U}+O(1)$, as per the requirements of the theorem, then writing
\begin{equation}
U= u^2
\end{equation}
will produce metric elements $G_{XX}$ and $G_{uu}$ which are regular and fall into the first of the two confining cases: an end of space, where the metric elements reach a minimum, non-negative and finite value. Thus, the distinction between two cases is in fact a little illusory as we have already mentioned. However, as was the case in the example we saw, performing this operation typically produces vanishing metric elements as a consequence, particularly in the shape of conical subspaces.

Thus, secondly, we must find a good procedure to unwrap cone-like structures in the metric, where compact bases shrink to zero size as we approach the end of space. In detail, we wish to write the cone-like metric, similar to a set of polar coordinates, in a more Cartesian way, as follows: if $S_n$ is some compact $n-$dimensional space, we want to rewrite the metric in terms of $n+1$ pseudo-Cartesian coordinates $Z_i$ such that
\begin{equation}
ds^2=A(u)dudu +  u^2 B(u) dS_n^2 = C\left( \sum_{i=1}^{n+1} Z_i^2\right)  \left(\sum_i^{n+1} dZ_idZ_i \right)
\label{arbmetric}
\end{equation}
and $C$ is non-vanishing around the point $Z_1=\dots Z_{n+1}=0$. We would then like to observe that $C$ has a minimum around $Z\cdot Z=0$, this would lead very generically to an identical interaction term that we have found in the specific example. This makes intuitive sense: the radial warp factor is constrained by the theorem to have a minimum around this point, this fact should transpire whichever coordinate system we care to use.

 It happens that this type of transformation is generically possible: in Lorentzian signature this is tantamount to the Kruskal-Szekeres change of coordinates. In particular it goes through an intermediate stage where one has to compute the Tortoise coordinate: with our definitions, this is the following function
\begin{equation}
F(u)=\int_0^u dx\, \frac{1}{x}\left( \sqrt{\frac{A(x)}{B(x)}} - 1\right) 
\end{equation}
While the total procedure is exact, in theory, in practice we would like workable metric elements out of it, that is, we'd like for this integral to be expressible analytically in terms of known functions. This can prove very difficult, and even when possible it can yield very unwieldy results. Thankfully, we already assumed that we would at some stage focus on the details of the metric close to the end of space: we only need to do this procedure locally around the surface of interest, as we did in the worked example.

Now, precisely due to the confinement theorem stated earlier, we know what the approximate geometry of the space near the IR surface looks like. After removing the horizon-like singularity, the theorem states that
\begin{equation}
	A(u)=a_0 + a_2 u^2 + \dots, 
\end{equation}
We mentioned that there is no reason to have $a_2$ vanish, generically one would expect a full set of coefficients in the expansion. $a_0$ needs to be positive by the requirements of the theorem. Furthermore, if the periods of the compact variables have been chosen correctly
\begin{equation}
	B(u)= a_0 + b_2 u^2 + \dots
\end{equation}
This function cannot have a linear term, it also needs to have a minimum at $0$: if it did, by a Weyl transformation we could shunt this behaviour as $u\rightarrow0$ onto the $G_{uu}$ metric element, which would then violate the theorem. Similarly, we don't expect $b_2$ to vanish, or to be equal to $a_2$, which would make $\frac{A}{B}$ of artificially higher order.

To first order, then, the Tortoise coordinate is invariably of the form 
\begin{equation}
	F(u)=  k u^2 +\cdots 
\end{equation}
where $k$ is determined from $a_0,\,a_2,\,b_2$. It is then sufficient to perform the following change of variables to unwrap the cone:
\begin{equation}
Z_i=u \exp(k u^2) w_i
\end{equation}
where the $w_i$ are a Cartesian parametrisation of the base space $S_n$ of unit radius, so that we have
\begin{equation}
	Z\cdot Z= u^2 \exp(2ku^2),
\end{equation}
a relationship which can easily be inverted to find $u$ as a function of $Z$.

 This ensure any conformally flat metric on the $Z_i$ maps to a conformally-equivalent metric to the one we seek, i.e. Eq.(\ref{arbmetric}). It is then simply a matter of finding the right conformal factor to get the exact result: in practice, one needs to find
\begin{equation}
	C(Z\cdot Z)= C(u^2 \exp(2ku^2)) = \exp(2ku^2),
\end{equation}
yet again $C$ does not need to be computed exactly, we only need to find its  approximate behaviour near $Z\cdot Z\rightarrow 0$.

By this definition, $C$ also has a minimum at $Z\cdot Z$=0, which confirms intuitive sense. Thus, in either of the two main cases, this entire procedure leads to a metric (and therefore a worldsheet effective theory) which takes the following form:
\begin{equation}
ds^2= dX^\mu dX_\mu + dZ_i dZ_i + \lambda Z\cdot Z dX^\mu dX_\mu + \cdots
\end{equation}
where $\lambda$ is a small parameter which controls the expansion of the metric. 

This shows that barring exceptional cases where the metric has an abnormal amount of flatness around the end of space, we will systematically get an interaction term of the form we derived in the practical example. 

For another worked example with a different type of conical singularity, the case of the Klebanov-Strassler\cite{Klebanov:2000hb} metric was treated in our original paper. The background is regular, there is no horizon, and its extra dimensions form a space broadly isomorphic to a cone whose base is made up of two spheres, $S_2\times S_3$.  Towards the end of space, the $S_3$-based cone caps out, it remains at finite size, while the $S_2$-based cone shrinks to zero size. Introducing three pseudo-Cartesian coordinates, this space can be unwrapped and made into a regular worldsheet theory which is structurally identical to the one we saw: the metric coefficients do not vanish abnormally, generating an interaction term exactly identical in form up to numerical coefficients.

While this is certainly a good consistency check, a more subtle one needs to be verified in higher-point amplitudes.

\subsection{Higher-point amplitudes}

We have mentioned our hypothesis is that the departure from the Veneziano regime can be entirely encapsulated in the Regge function $\alpha(s)$, which is linear for high values of its argument. Strictly speaking, it is something weaker that we are searching for: we seek extra terms that occur in the 4-point amplitude in the Regge regime, which we then interpret as stemming from a single function.

With this in mind it is then possible to extend our analysis to higher $n-$point amplitudes. The higher-order variants of the Veneziano amplitude are all defined generically in an integral form, the Koba-Nielsen integral.  In the case of the 4-point function the integral is easily expressed in terms of common functions, but much can be done using the general formulation:

\begin{equation}
	\mathcal{A}_n(k_i)=\int_{\mathbb{R}^n}\left(\prod_{i=1}^n dz_i\right) \left(\prod_{i<j} |z_i-z_j|^{k_i\cdot k_j} \right)  
\end{equation}

It is worth noting that the integral expressed above is written in its most stringy variant, and also its most indefinite. The integration volume has many infinite subspaces on which the integrand is constant, leading to divergences, much needs to be done to fix the exact value of the amplitude, in particular to define an integration measure that is sensible. An equivalent (and completely regular) way of writing the amplitude is given by the following expression
\begin{equation}
\int_{0}^1\cdots\int_{0}^1 \left( \prod\limits_{i=2}^{N-2} du_i u_i^{s_{1,i}-1}\left(1-u_i\right)^{ -s_{i,i+1}-1}  \right)\left(  \prod\limits_{i=2}^{N-3}\prod\limits_{j=i+2}^{N-1} \left(1-x_{ij} \right)^{s_{i,j}}\right) 
\end{equation}
where $x_{ij}=u_i u_{i+1} \dots u_j$ and $s_{i,j}=2p_i\cdot p_j$ are the various Mandelstam parameters in question. Although this form is suggestive of string scattering, one can postulate it in pure field theory from very general considerations of pole structure and Reggeon physics\cite{Bardakci:1969cs}.

Now, there exists a consistent picture for an equivalent multi-Regge limit for any number of operator insertions. In addition, a very straightforward modification of our worldsheet set-up is the only required change to treat the higher $n-$point case: most of the groundwork remains the same, the only difference is in the shape of the externally imposed current $J_0$ and thus the form of the 1-leg vertex. Nothing about the diagrammatic expansion changes, the same diagram remains the principal leading order contribution, immediately producing modifications to the Koba-Nielsen integrals. With the notation defined above, the aforementioned limit of the amplitude is to take the following operation:
\begin{align}
s_{i,i+1}\gg 1\,\,\, \,\,\, s_{1i}=\text{const.} \,\,\, , \,\,\, i=2\dots N-2 \nonumber \\
\frac{s_{i,i+1}s_{i+1,i+2}}{s_{i,i+2}}=-k_i=\text{const.}
\end{align}.

A full derivation of the multi-Regge formula will not be shown, but it is included in our original paper\cite{Armoni:2016llq}. The result of taking the aforementioned limit is the following
\begin{equation}
\left( \prod\limits_{i=2}^{N-2}\left(  s_{i,i+1}\right) ^{-s_{1,i}}\right) \times G\label{multiregge}
\end{equation}
where $G$ is an integral depending only on the constant parameters $k_i$ defined in the rules of the limit. 

 Including the corrections coming from the worldsheet computation, we get a modified Regge limit: from the very factorised form of the multi-Regge limit above we get that for all $i=2\dots N-2$ each of these leading terms in Eq.(\ref{multiregge}) is modified to
\begin{equation}
\left( s_{i,i+1}\right) ^{\left(-s_{1,i}\left(1-\rho^2 \log^2\left(s_{i,i+1}\right)  \right) \right) }
\end{equation}
for every $i$. This is the exact same form that suggested the corrected Regge function in the 4-point case, across all Regge channels of the higher point amplitude, this is encouraging, we are consistently recovering the same effect across multiple sources.

\section{Conclusions}

Understanding the true picture of meson scattering is an arduous task, which we hope to have shed some light on. The notions we have developed cement the argument that the Veneziano amplitude, while a fantastically useful tool, is not the final answer: it is the byproduct of strong coupling in an exclusive fashion, it bears no trace of weaker regimes inside of it. The quest to correct it, to form a more realistic picture of hadron scattering, is therefore of crucial importance.

As a summary of our findings, the broad procedure described in this paper is a novel way of exploiting the features of confining holographic backgrounds, which are receiving a great deal of interest recently. We use them in a very mathematical sense: we lean on their geometric features in order to generate for us the behaviour of a set of Wilson loops. Far from postulating that our world is sitting on the boundary of a higher-dimensional space, we merely use these backgrounds as a convenient mathematical tool, and an extremely natural one to use in the field theory set-up we described: the worldline formalism. We have put a deliberate amount of emphasis on showing the generic nature of this specific application of the concept, so we hope it can be of use in other settings.

It would be of great interest to test the limits of this kind of picture, since we have described only one possible application thereof, computing meson amplitudes. Representing functional determinants,  or more general sets of operator expectation values, as collections of Wilson Operators is very generically possible in practical examples of field theories, our holographic prescription then can be inserted in such computations in order to evaluate the Wilson Loop expectation values.

Foreseeably, one could also attempt to connect these ideas with Lattice Field Theory: the approach we have taken is very similar in spirit to the Strong Coupling and Hopping expansions, which in another work\cite{Armoni:2017dcr} we have shown is not at all accidental. We can hope that this correspondence can open up new ideas in either field.

\end{document}